\documentclass[12pt]{article}
\usepackage{amssymb,hyperref}
\usepackage[english]{babel}

\def\noi{\noindent}
\def\nqq{\hspace{-2em}}

\def\barr{\left(\begin{array}}
\def\earr{\end{array}\right)}
\def\beq#1{\begin{equation}\label{#1}}
\def\eeq{\end{equation}}
\def\ber#1{\begin{eqnarray}\label{#1} &&\nqq}
\def\eer{\end{eqnarray}}

\newcommand{\bear}[1]{\begin{eqnarray}\label{#1}}
\newcommand{\ear}{\end{eqnarray}}

\catcode`\@=11 \@addtoreset{equation}{section}\catcode`\@=12
\newcommand{\N}{ {\mathbb N} }
\newcommand{\R}{ {\mathbb R} }

\newcommand{\fnm}{\footnotemark}
\newcommand{\fnt}{\footnotetext}

\begin{document}

 \vspace{15pt}

 \begin{center}
 \large\bf

 On multidimensional analogs of Melvin's solution
 for classical series of Lie  algebras

 \vspace{15pt}

 \normalsize\bf
        A. A. Golubtsova\fnm[1]\fnt[1]{siedhe@gmail.com}$^{, b}$
        and   V. D. Ivashchuk\fnm[2]\fnt[1]{ivashchuk@mail.ru}$^{, a, b}$

 \vspace{7pt}

 \it (a) \ \ \ Center for Gravitation and Fundamental
 Metrology,  VNIIMS, 46 Ozyornaya Str., Moscow 119361, Russia  \\

 (b) \  Institute of Gravitation and Cosmology,
 Peoples' Friendship University of Russia,
 6 Miklukho-Maklaya Str.,  Moscow 117198, Russia \\

 \end{center}
 \vspace{15pt}

 \small\noi

 \begin{abstract}
 A multidimensional generalization of Melvin's
 solution for an arbitrary simple Lie algebra $\cal G$
 is presented. The gravitational model contains $n$ 2-forms and
 $l \geq n$ scalar fields, where $n$ is the rank of $\cal G$.
 The solution is governed by a set of $n$ functions obeying
 $n$ ordinary differential equations with certain boundary conditions. 
 It was conjectured earlier that these functions should
 be polynomials (the so-called fluxbrane polynomials). A program
 (in Maple)  for calculating of these polynomials for
 classical series of Lie  algebras is suggested (see Appendix).
 The polynomials corresponding to the Lie algebra $D_4$ are obtained.
 It is conjectured that the polynomials for  $A_n$-,$B_n$- and $C_n$-
 series  may be obtained  from   polynomials for $D_{n+1}$-series
 by using certain reduction formulas.

 \end{abstract}

 \section{Introduction}

  In this paper we deal with a special multidimensional generalization of
  the well-known Melvin solution \cite{Melv}. This generalized solution is related
  a simple Lie algebra and is a special case of the so-called generalized fluxbrane 
  solutions from \cite{Iflux}. For fluxbrane solutions see \cite{Iflux},
  \cite{GR}-\cite{GIM} and references therein. For more general classes of solutions 
  see also \cite{IK,IMtop}.

  We remind the reader that the generalized  fluxbrane solutions are governed by functions
  $H_s(z) > 0$ defined on the interval $(0, +\infty)$ and obeying the non-linear differential equations
  \beq{1.1}
  \frac{d}{dz} \left( \frac{ z}{H_s} \frac{d}{dz} H_s \right) =
   P_s \prod_{s' = 1}^{n}  H_{s'}^{- A_{s s'}},
  \eeq
 with  the following boundary conditions:
 \beq{1.2}
   H_{s}(+ 0) = 1,
 \eeq
 $s = 1,...,n$, where  $P_s > 0$  for all $s$.
 Parameters  $P_s$ are proportional to brane charge density squared
 $Q_s^2$ and $z = \rho^2$,  where $\rho$ is a radial parameter. The boundary condition
 (\ref{1.2}) guarantees the absence of singularity (in the metric) for $\rho =  +0$.

 Here  we assume that $(A_{s s'})$ is a Cartan matrix for some finite dimensional
 simple Lie algebra $\cal G$ of rank $n$ ($A_{ss} = 2$ for all $s$).
 According to a conjecture  suggested in \cite{Iflux}, the
 solutions to Eqs. (\ref{1.1}), (\ref{1.2}) governed by the Cartan matrix $(A_{s s'})$
 are  polynomials:  $H_{s}(z) = 1 + \sum_{k = 1}^{n_s} P_s^{(k)} z^k$,
  where $P_s^{(k)}$ are constants ($P_s^{(1)} = P_s$). Here
 $P_s^{(n_s)} \neq 0$  and $n_s = 2 \sum_{s' =1}^{n} A^{s s'}$
 where  $(A^{s s'}) = (A_{s s'})^{-1}$.
 Integers $n_s$ are components  of a twice dual
 Weyl vector in the basis of simple co-roots \cite{FS}.
 It was pointed in \cite{Iflux} that the conjecture on polynomial
 structure of $H_s$ may be proved for $A_n$ and $C_n$ Lie algebras along a line as it
 was done for black-brane polynomials from \cite{IMb1} (see also \cite{IMtop}).
 It should be also noted that
 the set of polynomials $H_s$ defines a special solution to  open Toda chain
 equations \cite{K,OP} corresponding to  simple Lie algebra $\cal G$.

\section{The solution}

We consider a  model governed by the action
 \beq{2.1}
  S=\int d^Dx \sqrt{|g|} \biggl \{R[g]-
  h_{\alpha\beta}g^{MN}\partial_M\varphi^{\alpha}\partial_N\varphi^{\beta}-\frac{1}{2}
  \sum_{s =1}^{n}\exp[2\lambda_s(\varphi)](F^s)^2 \biggr \}
 \eeq
where $g=g_{MN}(x)dx^M\otimes dx^N$ is a metric,
 $\varphi=(\varphi^\alpha)\in\R^l$ is a vector of scalar fields,
 $(h_{\alpha\beta})$ is a  constant symmetric non-degenerate
 $l\times l$ matrix $(l\in \N)$,    $ F^s =    dA^s
          =  \frac{1}{2} F^s_{M N}  dz^{M} \wedge  dz^{N}$
 is a $2$-form,  $\lambda_s$ is a 1-form on $\R^l$:
 $\lambda_s(\varphi)=\lambda_{s \alpha}\varphi^\alpha$,
 $s = 1,..., n$; $\alpha=1,\dots,l$.
 In (\ref{2.1}),
 we denote $|g| =   |\det (g_{MN})|$, $(F^s)^2  =
  F^s_{M_1 M_{2}} F^s_{N_1 N_{2}}  g^{M_1 N_1} g^{M_{2} N_{2}}$, $s = 1,..., n$.

 Let us consider a family of exact
solutions to the field equations corresponding to the action
(\ref{2.1}) and depending on one variable $\rho$. These solutions
are defined on the manifold
 \beq{2.2}
  M = (0, + \infty)  \times M_1 \times M_2,
 \eeq
 where $M_1$ is a one-dimensional manifold (say $S^1$ or $\R$) and
 $M_2$ is a (D-2)-dimensional Ricci-flat manifold. The solution
 reads
 \bear{3.30}
  g= \Bigl(\prod_{s = 1}^{n} H_s^{2 h_s /(D-2)} \Bigr)
  \biggl\{ w d\rho \otimes d \rho  +
  \Bigl(\prod_{s = 1}^{n} H_s^{-2 h_s} \Bigr) \rho^2 d\phi \otimes d\phi +
    g^2  \biggr\},
 \\  \label{3.31}
  \exp(\varphi^\alpha)=
  \prod_{s = 1}^{n} H_s^{h_s  \lambda_{s}^\alpha},
 \\  \label{3.32a}
  F^s= - Q_s \left( \prod_{s' = 1}^{n}  H_{s'}^{- A_{s
  s'}} \right) \rho d\rho \wedge d \phi,
  \ear
 $s = 1,..., n$, where $w = \pm 1$, $g^1 = d\phi \otimes d\phi$ is a
  metric on $M_1$ and $g^2$ is a  Ricci-flat metric on
 $M_{2}$.

 The functions $H_s(z) > 0$, $z = \rho^2$, obey the equations
(\ref{1.1}) with the boundary conditions (\ref{1.2}) and
 \beq{2.21}
  P_s =  \frac{1}{4} K_s Q_s^2.
 \eeq
 The parameters  $h_s$  satisfy the relations
  \beq{2.16}
  h_s = K_s^{-1}, \qquad  K_s = B_{s s} > 0,
  \eeq
 where
 \beq{2.17}
  B_{ss'} \equiv
  1 +\frac{1}{2-D}+  \lambda_{s \alpha} \lambda_{s' \beta}   h^{\alpha\beta},
  \eeq
 $s, s' = 1,..., n$, with $(h^{\alpha\beta})=(h_{\alpha\beta})^{-1}$.
 Here
 $\lambda_{s}^{\alpha} = h^{\alpha\beta}  \lambda_{s \beta}$
 and
 \beq{2.18}
  (A_{ss'}) = \left( 2 B_{s s'}/B_{s' s'} \right)
 \eeq
  is the Cartan matrix for a simple Lie algebra $\cal G$ of rank $n$.

It may be shown that if the matrix $(h_{\alpha\beta})$
 has an Euclidean signature and  $l \geq n$, there exists a set of co-vectors
 $\lambda_1, ..., \lambda_n$  obeying (\ref{2.18}).
 Thus the solution is valid at least when $l \geq n$
 and the matrix $(h_{\alpha\beta})$ is a positive-definite.

The solution under consideration may be verified just by 
substitution into the equations of motion corresponding to
 (\ref{2.1}). It may be also obtained as a special case of the
 fluxbrane (for $w = +1$,  $M_1 = S^1$) and $S$-brane
 ($w = -1$) solutions from \cite{Iflux} and \cite{GIM}, respectively.

 If $w = +1$ and the (Ricci-flat) metric $g_2$ has a
 pseudo-Euclidean signature, we get a multidimensional 
 generalization of the Melvin's solution \cite{Melv}. 
 Recall that the  Melvin's solution corresponds to
 $n = 1$, $M_1 = S^1$ ($0 < \phi <  2 \pi$),  $M_2 = \R^2$,
 $g_2 = -  dt \otimes dt + d \xi \otimes d \xi$
 and ${\cal G} = A_1$. For $w = -1$ and
 $g_2$ of Euclidean signature we obtain a cosmological solution
 with a horizon (as $\rho = + 0$) if $M_1 = \R$ ($ - \infty < \phi < + \infty$).

\section{Fluxbrane polynomials}

 Here we present polynomials corresponding to the Lie algebra $D_4 = {\rm so} (8)$.
 These polynomials were obtained using a program written in Maple.
 The program is given in the Appendix
 (it was described in \cite{GI}). \\

 For the Lie algebra $D_4$ we find the following set of polynomials
 \bear{D.1}
  H_{1} = 1 + P_1 z + \frac14 P_1 P_2 z^{2} + \Bigl( \frac{1}{36} P_1 P_2 P_3
  + \frac{1}{36} P_1 P_2 P_4 \Bigr) z^{3} + \frac{1}{144} P_1 P_2 P_3 P_4 z^{4}
  \\ \nonumber
  + \frac{1}{3600} P_1 P_2^{2} P_3 P_4 z^{5} + \frac{1}{129600}P_1^{2} P_2^{2} P_3 P_4 z^{6},\\
\label{D.2}
 H_{2} = 1 + P_2 z + \Bigl( \frac14 P_1 P_2 + \frac14 P_2 P_3 + \frac14 P_2 P_4 \Bigr) z^{2}
 + \Bigl( \frac19 P_1 P_2 P_3 + \frac19 P_1 P_2 P_4
 \\ \nonumber
  + \frac19 P_2 P_3 P_4 \Bigr) z^{3} + \Bigl( \frac{1}{144} P_1 P_2^{2} P_3
   + \frac{1}{144} P_1 P_2^{2} P_4 + \frac{1}{144} P_2^{2} P_3 P_4
    + \frac{1}{16} P_1 P_2 P_3 P_4 \Bigr) z^{4}
   \\ \nonumber
  + \frac{7}{600} P_1 P_2^{2} P_3 P_4 z^{5}
   + \Bigl( \frac{1}{1600} P_1 P_2^{3} P_3 P_4 +
   \frac{1}{5184} P_1 P_2^{2} P_3^{2} P_4 + \frac{1}{5184} P_1^{2} P_2^{2} P_3 P_4
    \\ \nonumber
   + \frac{1}{5184} P_1 P_2^{2} P_3 P_4^{2} \Bigr) z^{6}
    + \Bigl( \frac{1}{32400} P_1^{2} P_2^{3} P_3 P_4
    + \frac{1}{32400} P_1 P_2^{3} P_3 P_4^{2}
     + \frac{1}{32400} P_1 P_2^{3} P_3^{2} P_4 \Bigr) z^{7} \\ \nonumber
   + \Bigl( \frac{1}{518400} P_1^{2} P_2^{3} P_3 P_4^{2}
    + \frac{1}{518400} P_1^{2} P_2^{3} P_3^{2} P_4
    + \frac{1}{518400} P_1 P_2^{3} P_3^{2} P_4^{2} \Bigr) z^{8}
     \\ \nonumber
    + \frac{1}{4665600} P_1^{2} P_2^{3} P_3^{2} P_4^{2} z^{9}
     + \frac{1}{46656000} P_1^{2} P_2^{4} P_3^{2} P_4^{2} z^{10},\\
 \label{D.3}
 H_{3} = 1 + P_3 z + \frac14 P_2 P_3 z^{2}
 + \Bigl( \frac{1}{36} P_1 P_2 P_3 + \frac{1}{36} P_2 P_3 P_4 \Bigr) z^{3}
 + \frac{1}{144} P_1 P_2 P_3 P_4 z^{4}
  \\ \nonumber
 + \frac{1}{3600} P_1 P_2^{2} P_3 P_4 z^{5} + \frac{1}{129600} P_1 P_2^{2} P_3^{2} P_4 z^{6},\\
 \label{D.4}
 H_{4} = 1 + P_4 z + \frac14 P_2 P_4 z^{2} +
 \Bigl( \frac{1}{36} P_1 P_2 P_4 + \frac{1}{36} P_2 P_3 P_4 \Bigr) z^{3} +
 \frac{1}{144} P_1 P_2 P_3 P_4 z^{4}
  \\ \nonumber
  + \frac{1}{3600} P_1 P_2^{2} P_3 P_4 z^{5} + \frac{1}{129600} P_1 P_2^{2} P_3 P_4^{2} z^{6}.
  \ear

  Setting $P_4 = +0$, we get  a triple of polynomials
  $(H_1$, $H_2$, $H_3)$ for the Lie algebras $A_3 = {\rm sl}(4)$.
  For $P_3 = P_4$ we obtain a set of polynomials
  $(H_1$, $H_2$, $H_3)$ for the Lie algebras $B_3 = {\rm so}(7)$.
  The $A_3$-polynomials for $P_1 = P_3$ give us a pair of
  $C_2 = {\rm sp}(2)$-polynomials $(H_1$, $H_2)$ \cite{GIM}.
  (For $P_1 = P_3 = P_4$ we obtain $G_2$-polynomials from \cite{GIM}.)

  This prescription could be generalized to higher ranks
  by using the  following {\bf Conjecture}:
   i) the set of $A_{n}$-polynomials is given by the first $n$
   polynomials for the Lie algebra $D_{n+1}$ when $P_{n+1} = +0$;
   ii) the set of $B_{n}$-polynomials coincides with the first $n$
   polynomials for the Lie algebra $D_{n+1}$ when $P_n = P_{n+1}$;
   iii) the set of $C_{m +1 }$-polynomials coincides with the first $m +1$
   polynomials for the Lie algebra $A_{2m+1}$ when
    the following relations are imposed: $P_1 = P_{2m+1}$, $P_2 = P_{2m}$,
    ...,   $P_m = P_{m+2}$. An analytical proof of this conjecture will be a subject of
    a separate work. At the moment, one can verify these  relations just by using the program
    from the Appendix.

  \section{\bf Conclusions}

  We have presented a multidimensional generalization of the Melvin's
  solution for an arbitrary simple Lie algebra $\cal G$.
  The solution is governed by a set of $n$  fluxbrane polynomials.
  We have written a program for calculating of these polynomials
  for the classical series of Lie algebras  (see the Appendix).
  The set of polynomials corresponding to the  Lie algebra $D_4$
  is obtained.  The polynomials considered above define special solutions to open Toda chain
  equations corresponding to simple Lie algebras that may be of interest for
  certain applications of Toda chains.

  We have conjectured (without proof) certain relations  between
  polynomials belonging to different series of classical Lie algebras.
  These relations tells us that the most important is
  the calculation of $D_{n+1}$-polynomials, since all other polynomials
  (e.g. $A_n$-,$B_n$- and $C_n$-ones) may be obtained from the 
  polynomials for $D_{n+1}$-series by using certain reduction formulas.
  A calculation of polynomials for exceptional Lie algebras (i.e. $G_2$,
  $F_4$, $E_6$, $E_7$, $E_8$) will be described in a separate publication.

  As it was noted, the solution under consideration for $w = -1$
  is a cosmological one. The study of this and/or some other
  similar cosmological solution governed by fluxbrane polynomials
  (e.g. in connection with the problem of acceleration)
  will be a subject of a forthcoming publications.

  \begin{center}
  {\bf Acknowledgments}
  \end{center}

 This work was supported in part by the Russian Foundation for
 Basic Research grant  Nr. $07-02-13624-ofi_{ts}$ and
 by a grant of People Friendship University (NPK MU).

   \begin{center}
  {\bf Appendix}
   \end{center}

   In this Appendix we present a program written for calculating fluxbrane polynomials
   for four classical series of simple Lie algebras: $A_n$, $B_n$, $C_n$ and $D_n$.
   A description of this program is given  in \cite{GI}.

{\scriptsize
\begin{verbatim}
> with(LinearAlgebra):
> with(PolynomialTools):
> S:=3:
> A:=Matrix(S,S):
> AlgLie:=proc(algn,S,CartA:=Matrix(S,S))
     local i,mu,nu;
     i := 0; mu := 0; nu:=0;
     mu := S-1;
     nu:=S-2;
     for i to S do
          CartA[i, i] := 2
     end do;
     if (S>=1) and algn=an then
         for i from 1 to S-1 do
              CartA[i, i+1] := -1
         end do;
         for i from 1 to S-1 do
               CartA[i+1, i] := CartA[i, i+1]
         end do;
      end if;
      if (S >= 4) and algn = dn then
          for i from 1 to S-2 do
                CartA[i, i+1] := -1
          end do;
          for i from 1 to S-2 do
               CartA[i+1, i] := CartA[i, i+1]
          end do;
          CartA[S, nu] := CartA[mu, nu];
          CartA[nu, S] := CartA[S, nu]
      end if;
      if (S>=2) and algn=cn then
          for i from 1 to (S - 1) do
                CartA[i,i+1]:=-1
          end do;
          for i from 1 to (S - 2) do
               CartA[i+1,i]:=CartA[i,i+1]
          end do;
          CartA[S,mu]:=-2
      end if;
      if  (S>=3) and algn=bn then
           for i from 1 to (S - 1) do
                 CartA[i+1,i]:=-1
           end do;
           for i from 1 to (S - 2) do
                 CartA[i,i+1]:=CartA[i+1,i]
           end do;
      end if;
           return CartA
      end proc:
> AlgLie(algn, S, A);
> n := Vector[row](1 .. S):
> A1 := MatrixInverse(A);
> for i to S do n[i] := 2*add(A1[i, j], j = 1 .. S) end do:
> maxel := max(convert(n, list)[]):
> P := array(1 .. S, 1 .. maxel):
> H := Vector[row](1 .. S):
> for i to S do H[i] := 1+add(P[i, k]*z^k, k = 1 .. n[i]) end do:
> for i to S do for j to S do a[i, j] := A[i, j] end do end do:
> for i to S do h[i] := H[i] end do:
> for i to S do for v to S do c[i, v] := h[v]^(-a[i, v]) end do end do:
> equal := Vector[row](1 .. S):
> for i to S do
  equal[i]:= diff(z*(diff(H[i],z))/H[i],z)-P[i,1]*(product(c[i,m], m = 1..S))
  end do:
> simequal := Vector[row](1 .. S):
> newequal := Vector[row](1 .. S):
> for i to S do simequal[i] := simplify(combine(value(equal[i]), power)) end do:
> for i to S do newequal[i] := numer(simequal[i]) end do:
> maxcoeff := Vector[row](1 .. S):
> for i to S do maxcoeff[i] := degree(newequal[i], z) end do:
> coefflist := table():
> for i to S do
    for c from 0 to maxcoeff[i] do
        coefflist[i, c] := coeff(newequal[i], z, c) = 0
    end do
  end do:
> Sys := convert(coefflist, list):
> sol := solve(Sys):
> trans := {seq(seq(P[i,j] = P[i,j], i = 1..S), j = 1..maxel)}:
> sol := simplify(map2(subs, trans, sol)):
> P1 := map2(subs, sol, evalm(P)):
> for i to S do H[i]:= 1+add(P1[i,k]*z^k, k = 1..n[i]) end do;
\end{verbatim}}

{\small

 \end{document}